\documentclass[aps,prl,groupedaddress,twocolumn,showpacs]{revtex4-1}
\usepackage[dvips]{graphicx}
\usepackage{dcolumn}
\usepackage{bm}
\usepackage{amssymb}
\usepackage{epstopdf}
\usepackage{amsmath}
\usepackage{color}
\usepackage[dvipdfmx]{hyperref}
\hypersetup{
 setpagesize = false,
 colorlinks = true,
 citecolor = blue,
 linkcolor = blue,
 urlcolor = blue,
 }
\begin{document}
\title{Achieving high-temperature ferromagnetism by means of magnetic ions dimerization}
\author{Panjun Feng$^{1}$}\email{These authors contributed equally to this work.}
\author{Shuo Zhang$^{1}$}\email{These authors contributed equally to this work.}
\author{Dapeng Liu$^{1}$}\email{These authors contributed equally to this work.}
\author{Miao Gao$^{2}$}
\author{Fengjie Ma$^{3}$}
\author{Xun-Wang Yan$^{1}$}\email{Corresponding author: yanxunwang@163.com}
\author{Z. Y. Xie$^{4}$}\email{Corresponding author: qingtaoxie@ruc.edu.cn}
\date{\today}
\affiliation{$^{1}$College of Physics and Engineering, Qufu Normal University, Qufu, Shandong 273165, China}
\affiliation{$^{2}$Department of Physics, School of Physical Science and Technology, Ningbo University, Zhejiang 315211, China}
\affiliation{$^{3}$The Center for Advanced Quantum Studies and Department of Physics, Beijing Normal University, Beijing 100875, China}
\affiliation{$^{4}$Department of Physics, Renmin University of China, Beijing 100872, China.}
\begin{abstract}
 Magnetic two-dimensional materials have potential application in next-generation electronic devices and have stimulated extensive interest in condensed matter physics and material fields. However, how to realize high-temperature ferromagnetism in two-dimensional materials remains a great challenge in physics.
 Herein, we propose an effective approach that the dimerization of magnetic ions in two-dimensional materials can enhance the exchange coupling and stabilize the ferromagnetism.
  Manganese carbonitride Mn$_2$N$_6$C$_6$ with a planar monolayer structure is taken as an example to clarify the method, in which two Mn atoms are gathered together to form a ferromagnetic dimer of Mn atoms and further these dimers are coupled together to form the overall ferromagnetism of the two-dimensional material.
  In Mn$_2$N$_6$C$_6$ monolayer, the high-temperature ferromagnetism with the Curie temperature of 272.3 K is determined by solving Heisenberg model using Monte Carlo simulations method.

\end{abstract}


\maketitle

\section{Introduction}

Owing to the reduced dimensionality and symmetry, two-dimensional (2D) materials possess a variety of unique qualities associated with exotic and fantastic physical phenomena,
which provide tremendous opportunities in both fundamental study and potential application research.
Since graphene was discovered in 2004 \cite{Novoselov2004}, 2D materials have attracted great attention from researchers in physics and material fields.
In experiments, various 2D materials have been fabricated, such as boron nitride \cite{Song2010}, silicene \cite{Lalmi2010}, borophene \cite{Mannix2015}, stanene \cite{Saxena2016}, transition metal dichalcogenides \cite{Coleman2011}, MXenes \cite{Naguib2011}, $etc$.
Meanwhile, theoretical researches have designed and predicted a variety of 2D compounds, such as Si$_2$BN \cite{Andriotis2016}, B$_3$C$_2$ \cite{Wang2017},
Fe$_2$Si \cite{Sun2017}, CoB$_6$ \cite{Tang2019}, and so on.
However, among the reported 2D materials, the proportion of magnetic 2D compounds is very small.

Ferromagnetic 2D material is a kind of highly promising material for the next-generation spintronic devices including giant magneto-resistance, spin valve, magnetic random-access memory and other spin logic devices, and the research on 2D ferromagnet is extremely vigorous in recent years.
In 2017, long-range ferromagnetic ordering was first discovered in CrI$_3$ monolayer by scanning magneto-optic Kerr microscopy and the Curie temperature is 45 kelvin.\cite{Huang2017}
Subsequently, ferromagnetic order in exfoliated Cr$_2$Ge$_2$Te$_6$ bilayer was observed,\cite{Gong2017} and the Curie temperature of 330 kelvin in VSe$_2$ monolayer was also reported. \cite{Bonilla2018}
Very recently, robust room-temperature ferromagnetism with $T_C$ up to 400 K has been realized in CoN$_4$-embedded graphene. \cite{Hu2021}
In theoretical studies, CrI$_3$ \cite{Zhang2015}, CrGeTe$_3$ \cite{Sivadas2015}, Fe$_2$Si \cite{Sun2017}, MnS$_2$ \cite{Guan2019}, Fe$_3$P \cite{Zheng2019}, CrMoS$_2$Br$_2$ \cite{Chen2020}, GdI$_2$ \cite{Wang2020}, CrN$_4$C$_2$ \cite{Liu2021b} monolayers are predicted to be 2D ferromagnet with high Curie temperature.

Although some progress has been made, the study on ferromagnetic 2D materials is still in the primary stage.
Curie temperature is a decisive physical parameter to the properties of ferromagnetic materials.
At present, Curie temperature much lower than room temperature is the main obstacle to the applications of 2D ferromagnets in magnetic device fabrication.
A natural and fundamental question is how to enhance the Curie temperature of a 2D ferromagnet.
 Multiple factors can influence the Curie temperature of 2D materials. Specifically,
 it depends on the exchange coupling between magnetic ions, the spin of each ion, and magnetic anisotropic energy, and the type of crystal lattice.
A few schemes including strain, carrier doping, adsorption, and proximity effect are employed to enhance the exchange interaction
 in the known 2D materials. \cite{Guo2020,Huang2020}
In these methods, the spin $S$ only refers to single atom spin and the coupling $J$ is only related two single atoms, rather than the clusters of magnetic atoms, which gives rise to small $S$ and $J$ and limits the realization of high critical temperature.

In this work, we propose a radically different strategy to enhance magnetic exchange energy.
First, two magnetic ions are gathered together to form a ferromagnetic dimer as one lattice point, which has a larger total spin and magnetic anisotropic energy than single magnetic ion.
 Then, a suitable distance between two dimers is chosen to ensure a strong magnetic exchange coupling.
In previous experiments, the dimers of transition metal atoms were embedded in N-doped graphene to synthesize the dual-atom catalyst \cite{Lu2019}, and even the structural unit of Mn trimer was also realized in recently synthesized MnSn multilayer\cite{Yuan2020}.
These clusters of two or three atoms are embedded in the two-dimensional monolayer and are small enough to be regarded as a lattice point on a two-dimensional lattice.
Based on the above idea and the experimental works,
we screen out the planar Mn$_2$N$_6$ as a basic unit, which is a Mn dimer and has total magnetic moments up to 7.3 $\mu_B$.
Then, the Mn$_2$N$_6$ moieties are embedded in graphene lattice to make up a hybridized monolayer with the distances among Mn$_2$N$_6$ moieties as short as possible.
After much trial and error, we design a 2D manganese carbonitride Mn$_2$N$_6$C$_6$ with the Curie temperature of 272 K, which is a fairly near-room-temperature predicted by the Heisenberg model \cite{Guo2020}.

\section{Computational Method}
The calculations are performed in VASP package, in which the plane wave pseudopotential method and the projector augmented-wave (PAW) pseudopotential with Perdew-Burke-Ernzerhof (PBE) functional \cite{PhysRevB.47.558, PhysRevB.54.11169, PhysRevLett.77.3865, PhysRevB.50.17953} are adopted, and Hubbard U is included to consider the correction of electron correlation within GGA + U method \cite{Cococcioni2005}.
The plane wave basis cutoff is 600 eV and the thresholds are 10$^{-5}$ eV and 0.01 eV/\AA ~ for total energy and force convergence.
The interlayer distance was set to 18 \AA~ and a mesh of $24\times 24\times 1$ k-points is used for the Brillouin zone integration.
The phonon calculations are carried out with the supercell method in the PHONOPY program,
and the real-space force constants of supercells were calculated using density-functional perturbation theory (DFPT) as implemented in VASP  \cite{Togo2015}.
The force convergence criterion (10$^{-5}$ eV/\AA) was used in structural optimization of the primitive cell before building the supercell.
In the ab initio molecular dynamics simulations,
the 2 $\times$ 2 $\times$ 1 supercells were employed and the temperature was kept at 1000 K for 5 ps with a time step of 1 fs in the canonical ensemble (NVT) \cite{Martyna1992}.
The temperature of phase transition in Mn$_2$N$_6$C$_6$ systems is evaluated by solving the Heisenberg model with Monte Carlo method, implemented in the Mcsolver package \cite{mcsolver,Liu2020}.
Monte Carlo simulations are performed with a 60 $\times$ 60 $\times$ 1 spin lattice in which the spins interact with each other. For one temperature value, the system of spins corresponds to one thermal equilibrium state, and the related statistics observables such as energy and magnetic moment are directly obtained and other physical quantities can be derived. The average moment and the specific heat capacity are expressed as $M = \frac{1}{N}|\sum_i \vec{M_i}|$ and $C_v = \frac{<E^2> - <E>^2}{k_BT^2}$.

\begin{figure}
\begin{center}
\includegraphics[width=7.5cm]{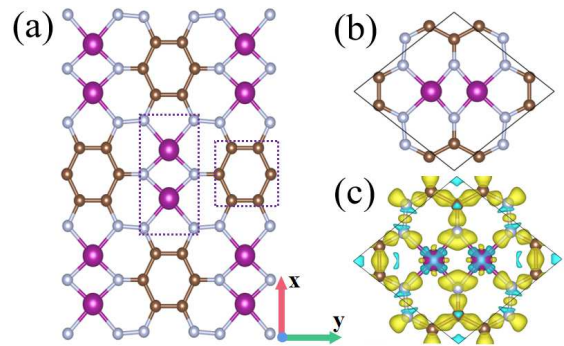}
\caption{(a) Atomic structure of Mn$_2$N$_6$C$_6$  monolayer containing Mn$_2$ dimer. (b) A unit cell of Mn$_2$N$_6$C$_6$. (c) Charge density difference relative to the superposition of atomic charge density. Yellow and blue surfaces show charge accumulation and depletion, respectively.
 } \label{structmodel}
\end{center}
\end{figure}

\section{Results and Discussion}

Fig. \ref{structmodel}(a) shows the crystal structure of Mn$_2$N$_6$C$_6$ monolayer, which consists of six-members carbon rings and Mn$_2$N$_6$ moieties, depicted with the dashed line rectangle. The two components are arranged alternately to form a planar layer, in which all nitrogen and carbon atoms are tricoordinated and metal atoms are tetracoordinated. The primitive cell of Mn$_2$N$_6$C$_6$ is displayed with a solid line rhombus and has the symmetry of plane group $Cmm$, shown in Fig. \ref{structmodel}(b). In Mn$_2$N$_6$ moiety, two MnN$_4$ units are fused together by two shared N atoms, which results in the dimerization of two adjacent Mn atoms.
Fig. \ref{structmodel}(c) shows the charge density difference relative to atomic charge density, in which yellow and blue surfaces depict the charge accumulation and depletion. Yellow surfaces between N and Mn atoms indicate the formation of N-Mn coordination bonds,
while three yellow pockets around N and C atoms reflect their $sp^2$ orbital hybridization and the N-N and N-C covalent bonds. So the framework made up of C and N atoms with $sp^2$ orbital hybridization is the main reason why the monolayer is stable, while the large $\pi$ bond stemming from their half-filled $p_z$ orbitals further strengthen the stability and make the layer keep in a plane.
For Mn atom, it is coordinated by four N atoms and localized at the center of N approximate square. Apart from Mn-N coordination bond, there exist $\pi$-$d$ conjugation between $\pi$ bond of N-C framework and Mn $d_{xz}$ and $d_{yz}$ orbitals, which leads to larger $\pi$ bond extending to the overall planar monolayer.
These bonding characters maintain the planarity and stability of the Mn$_2$N$_6$C$_6$ monolayer.

\begin{figure}[htbp]
\begin{center}
\includegraphics[width=8.0cm]{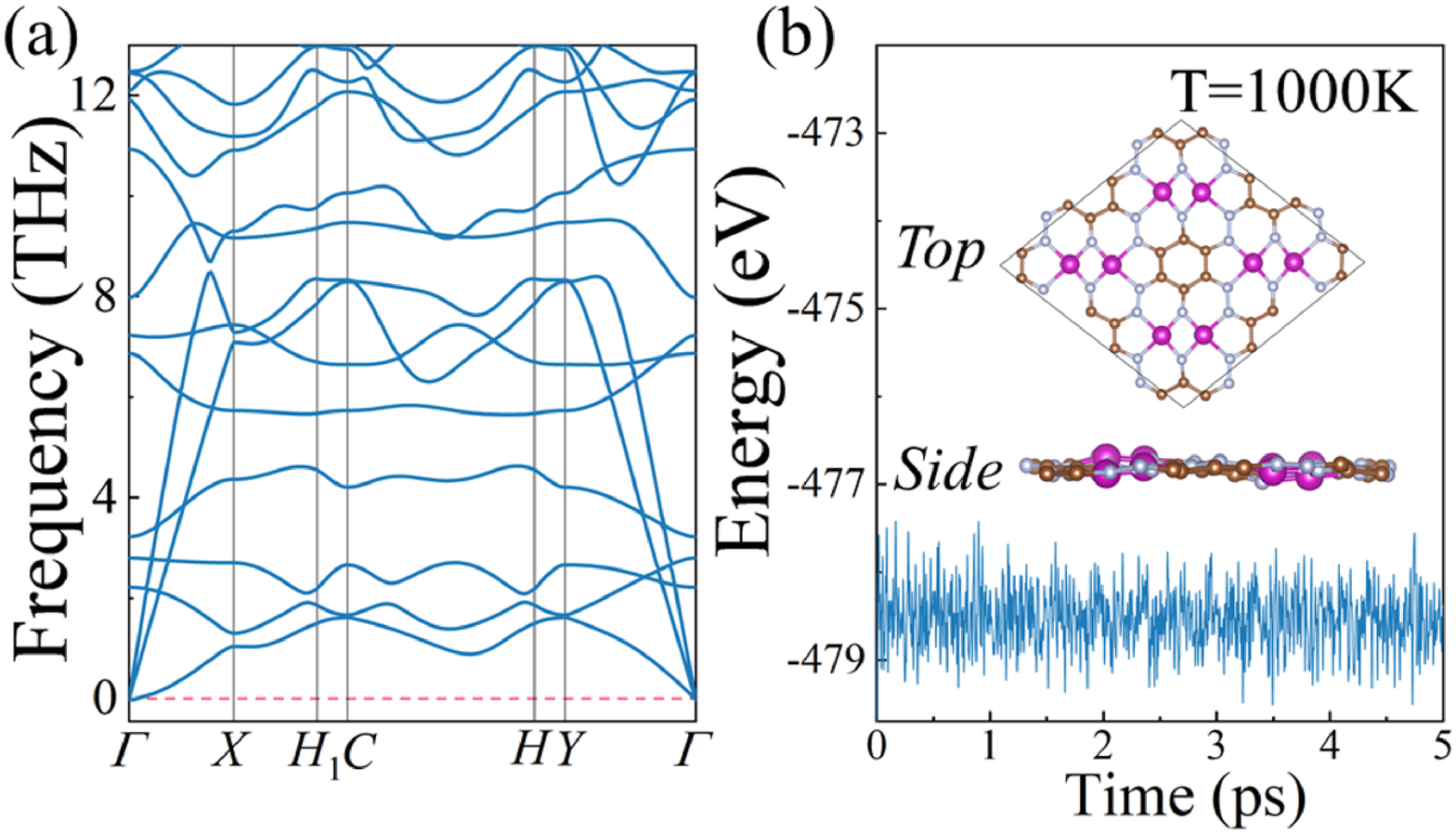}
\caption{(a) Phonon spectra of Mn$_2$N$_6$C$_6$ monolayer; (b) Total energy evolution of Mn$_2$N$_6$C$_6$ with respect to time in molecular dynamics simulation. The insets are the top and side views of final configurations for Mn$_2$N$_6$C$_6$ monolayer at 1000 K after 5 ps.
  } \label{phonon}
\end{center}
\end{figure}

To clarify the structural stability of Mn$_2$N$_6$C$_6$ monolayer,
we perform the calculations of formation energy, phonon spectra, molecular dynamics simulations.
The formation energy ${E}_{form}$ is defined by
$E_{form} = \frac{1}{14}*(E_{tot} - 2E_{Mn} - 3E_{N_2} - 6E_C)$ = 0.107 eV for Mn$_2$N$_6$C$_6$,
 in which $E_{tot}$, $E_{Mn}$, $E_{N_2}$, and $E_C$ are the total energy of Mn$_2$N$_6$C$_6$, bulk metal energy per atom, nitrogen molecule energy, graphene energy per atom, respectively.
We also compute the formation energy of $g$-C$_3$N$_4$ in terms of the above method and the value is 0.35 eV.
From the viewpoint of formation energy, the planar structure of Mn$_2$N$_6$C$_6$ is more stable than $g$-C$_3$N$_4$ that has already been synthesized in experiments \cite{Groenewolt2005}.
The phonon spectra of Mn$_2$N$_6$C$_6$ structure is calculated in ferromagnetic order, shown in Fig. \ref{phonon}(a).
No imaginary frequency mode occurs in the phonon spectra.
Moreover, we simulate the molecular dynamics behavior of Mn$_2$N$_6$C$_6$ at the temperature of 1000 K and the structure is well maintained in the whole evolution process of total energy in 5 ps.
The fluctuation of total potential energy of Mn$_2$N$_6$C$_6$ monolayer and the final configuration at the end of the simulation are presented in Fig. \ref{phonon}(b).
So, the dynamical and thermal stability of Mn$_2$N$_6$C$_6$ is verified.
In fact, the structure of Mn$_2$N$_6$C$_6$ is similar to the MnN$_4$-embedded graphene sheets, which have been synthesized in a great number and widely studied in previous theoretical and experimental works \cite{Fei2018,Zhao2019,Zhang2020,Hu2021,Liu2021,Liu2021a}.

\begin{figure}[htbp]
\begin{center}
\includegraphics[width=8.5cm]{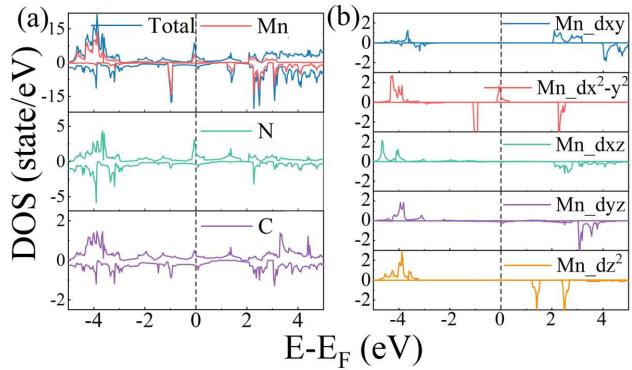}
\caption{(a) Total density of state of Mn$_2$N$_6$C$_6$ and the density of states projected on C, N, and Mn atomic orbitals. (b) Density of states projected on five 3$d$ suborbitals of Mn atom. Fermi energy is marked by the vertical dashed line.
  } \label{total-dos}
\end{center}
\end{figure}

The electronic structure of Mn$_2$N$_6$C$_6$ is studied by the GGA + U calculations, which is a more precise method than pure PBE functionals method to describe the electronic correlation of transition metal compounds. The advantage of GGA + U method is that Hubbard U has a clear physical significance, while the disadvantage is that Hubbard U value is difficult to determine because it is not identical for different compounds. To avoid its disadvantage, M. Cococcioni and S. de Gironcoli develop a self-consistent approach to estimate the U parameter \cite{Cococcioni2005}, and the method is contained in the example category in the Vaspwiki website. By this method, the Hubbard U is derived for the Mn$_2$N$_6$C$_6$ monolayer,
$ U=\chi ^{-1}-\chi _{0}^{-1}\approx \left({\frac {\partial N_{I}^{\rm {SCF}}}{\partial V_{I}}}\right)^{-1}-\left({\frac {\partial N_{I}^{\rm {NSCF}}}{\partial V_{I}}}\right)^{-1}={\frac {1}{0.13}}-{\frac {1}{0.81}}=4.88\;eV $.
Fig.\ref{total-dos}(a) shows the total density of states of Mn$_2$N$_6$C$_6$ monolayer and the projected density of states on C, N, and Mn atoms, which indicate that the electronic states near Fermi energy are mainly from the N and Mn atoms.
As shown in Fig. \ref{total-dos}(b), the $d_{z^2}$, $d_{xz}$, and $d_{yz}$ orbitals are fully spin-polarized and make a main contribution to the large moment, while the $d_{x^2-y^2}$ and $d_{xy}$ states have partial and very small contributions.
The moment of Mn atom reaches up to 3.6 $\mu_B$.
Since N atom is directly connected to Mn atom, its $2p$ states are affected by the polarized $d$ electrons and some spin polarization is induced.
In addition,
the N $s$, $p_x$, $p_y$ orbitals and Mn $d_{xy}$ orbital are distributed in the similar energy range (not shown), corresponding to the $sp^2$ hybridization and the N-Mn coordination bond, which make up the framework of Mn$_2$N$_6$C$_6$ monolayer.
N $p_z$, C $p_z$, Mn $d_{xz}$ and $d_{yz}$ states have the obvious overlap in their energy ranges, which is closely related to $\pi$ conjugation between N and C atoms and the $\pi-d$ conjugation between N and Mn atoms. The $\pi$ and  $\pi-d$ conjugations play a key role in maintaining the planarity of Mn$_2$N$_6$C$_6$ monolayer.

\begin{figure}[htbp]
\begin{center}
\includegraphics[width=8.0cm]{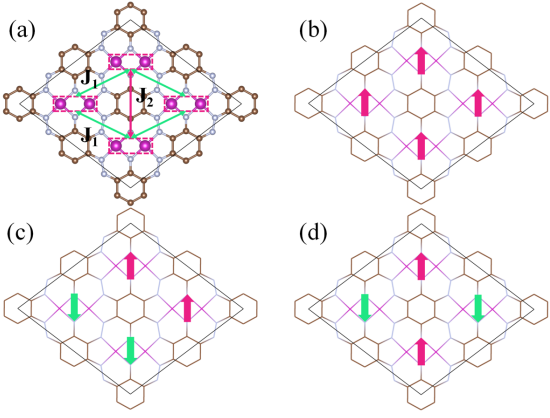}
\caption{(a) Top view of Mn$_2$N$_6$C$_6$ monolayer. $J_1$ and $J_2$ are the exchange interactions of two neighboring sites in distorted triangle lattice. (b) FM order; (c) AFM-I order; (d) AFM-II order. The solid line rhombus represents the magnetic unit cell. To exhibit the magnetic order more clearly, the atomic structure is displayed with the wire frame.
  } \label{order}
\end{center}
\end{figure}

As for the magnetism of Mn$_2$N$_6$C$_6$ layer, we first pay attention to the magnetic coupling between two Mn moments in a dimer in Mn$_2$N$_6$C$_6$ layer.
The two Mn moments can be aligned in the same or opposite direction. The energy for parallel alignment is 56.4 meV lower than that for antiparallel case,
which indicates that there is strong ferromagnetic interaction in the dimer.
The angle of Mn-N-Mn in Mn$_2$N$_6$C$_6$ is 81.7$^{\circ}$, close to 90$^{\circ}$.
According to the Goodenough-Kanamori-Anderson (GKA) rules \cite{Goodenough1955,Anderson1959}, when the angle of cation-anion-cation is about 90$^\circ$ the superexchange interaction favors ferromagnetic alignment.
Consequently, the dimer of Mn atom can be regarded as a lattice point whose moment and magnetic anisotropy energy are the sum of the anisotropy energy and magnetic moment of two Mn atoms.
Next, we demonstrate that the ferromagnetic order is the magnetic ground state of Mn$_2$N$_6$C$_6$ layer.
The ferromagnetic order (FM), antiferromagnetic order I (AFM-I), and antiferromagnetic order II (AFM-II) are sketched in Fig. \ref{order}.
There, the red and blue arrows mean the direction of total moment of a dimer.
For each formula unit cell, the energies of AFM-I and AFM-II states are 74.4 meV and 148.2 meV higher than FM energy.
We also carry out the HSE hybridized functional calculations \cite{Krukau2006} to examine the energy order of FM, AFM-I, and AFM-II states, and the ferromagnetic ground state of Mn$_2$N$_6$C$_6$ monolayer is confirmed.

On account of the existence of large local moments in Mn$_2$N$_6$C$_6$ monolayer, we use Ising model to describe the magnetic interactions among lattice points, which is the simplest model to show a magnetic phase transition \cite{Sun2017,Zhao2020,Guo2020}.
The Hamiltonian is shown below,
\begin{equation}
H = \sum_{<ij>}J_1 S_i S_j + \sum_{\ll ij^{\prime} \gg}J_2 S_i S_{j^{\prime}},
\end{equation}
where $j$ and $j^\prime$ denote the neighbors of $i$ site along different direction, and
$E_{FM}$, $E_{AFM-I}$, and $E_{AFM-II}$ are the energies of FM, AFM-I, and AFM-II orders per formula cell.
In terms of the formulas \ref{J1J2}, the exchange couplings $J_1$ and $J_2$ between two neighboring dimers are -37.04 meV/S$^2$ and -0.18 meV/S$^2$ \cite{Ma2008}.
\begin{equation}
\begin{aligned}\label{J1J2}
J_1 &= \frac{1}{4}(E_{FM} - E_{AFM-II}),  \\
J_2 &= \frac{1}{4}(E_{FM} + E_{AFM-II} - 2E_{AFM-I})
\end{aligned}
\end{equation}

\begin{figure}[htbp]
\begin{center}
\includegraphics[width=8.0cm]{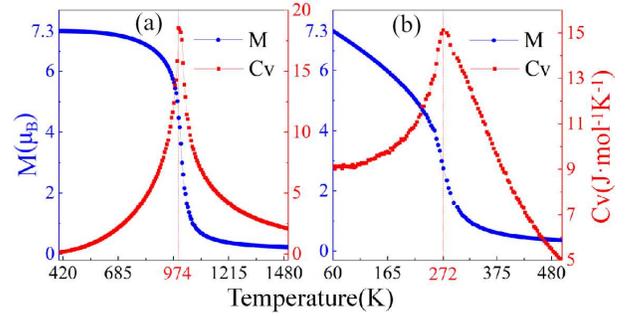}
\caption{The specific heat capacity $C_v$ and average magnetic moment $M$ as functions of temperature for the Ising model (a) and Heisenberg model (b).
The phase transition temperatures are marked by the vertical dashed lines.
  } \label{Tc}
\end{center}
\end{figure}

The spontaneous magnetization ($M$) and specific heat ($C_v$) with respect to temperature are plotted in Fig. \ref{Tc}(a),
and the Curie temperature of Mn$_2$N$_6$C$_6$ is determined with Monte Carlo method to be 974 K.
If we ignore $J_2$ and just focus on $J_1$, the magnetic interactions in Mn$_2$N$_6$C$_6$ layer can be regarded as a 2D square lattice.
Its Curie temperature can be estimated by the analytical solution of Ising model on square lattice, namely, $T_c = 2J_1/ln(1+\sqrt{2})$ = 2.269 $\times$ 37.04 $\times$ 11.6 = 974.8 K. The values of Curie temperature with two methods are consistent with each other.

Due to the small magnetic anisotropy in real 2D ferromagnetic materials, Curie temperature is usually overestimated by Ising model. Then, we turn to Heisenberg model, which is more precise model and has been successfully used to estimate the Curie temperature of CrI$_3$ and other synthesized ferromagnetic 2D materials \cite{Liu2016,Guo2020}.
For example, the calculated Curie temperature of 42 K is very close to the experimental value of 45 K in 2D ferromagnetic CrI$_3$ layer \cite{Zhang2021}.
The Hamiltonian of Heisenberg model is defined as
\begin{equation}
\label{Heisenberg}
H = \sum_{<ij\alpha>}J_{1\alpha}{S}_{i\alpha}{S}_{j\alpha} + \sum_{\ll ij^{\prime}\alpha \gg}J_{2\alpha}{S}_{i\alpha}{S}_{j^{\prime}\alpha} + A\sum_{i}(S_{iz})^2,
\end{equation}
in which the symbol $j$ and $j^\prime$ represent the neighboring sites of $i$ site in the distorted triangle lattice, and $\alpha$ is coordinate component $x$, $y$, or $z$. $A$ is the single-site magnetic anisotropic energy.
We perform the GGA + U + SOC calculations with the spin along (1 0 0), (0 1 0), and (0 0 1) axis for FM, AFM-I, and AFM-II magnetic orders, and the magnetic exchange interactions can be derived from these energy differences.
The exchange couplings $J_{1x}$, $J_{1y}$, $J_{1z}$, $J_{2x}$, $J_{2y}$, and $J_{2z}$ are -36.08, -36.10, -36.09, 0.07, 0.10, and 0.09 meV/S$^2$, respectively, while the single-site magnetic anisotropic energy is 0.765 meV/S$^2$ with easy axis normal to Mn$_2$N$_6$C$_6$ monolayer.
Fig. \ref{Tc}(b) displays the variation of magnetic moment and the specific heat with temperature for Mn$_2$N$_6$C$_6$ monolayer,
which gives the Curie temperature of 272 K.
If the superexchange couplings between two more distant sites are taken into account, the critical temperature is 270.9 K, which manifests the reliability of the above Curie temperature derived from $J_1$ and $J_2$.
The ferromagnetic critical temperature is much lower than 974 K predicted based on Ising model, indicating that Ising model significantly overestimate the Curie temperature of Mn$_2$N$_6$C$_6$ monolayer especially in the case of small magnetic anisotropy.
Yilv Guo $et ~al$ summarized the recent progress of ferromagnetic 2D materials in their review article \cite{Guo2020}, and the Curie temperatures predicted by Ising model and Heisenberg model were listed in Table 1 in Ref.\citenum{Guo2020}, in which the temperature from Ising model is overestimated.
On the other hand, among 44 kinds of ferromagnetic 2D materials listed in Ref \citenum{Guo2020}, the highest Curie temperature predicted based on Heisenberg model is 261 K, which indicates that Mn$_2$N$_6$C$_6$ monolayer with Curie temperature of 272 K is a near-room-temperature ferromagnetic 2D material.

\section{Conclusion}
In summary, by the first-principles calculations, we propose an effective approach to realize the high-temperature ferromagnetism in 2D materials. Namely, two magnetic ions are gathered together to form a ferromagnetic dimer, which acts as one lattice point and has twice moment of single magnetic atom.
Based on the idea, we design the Mn$_2$N$_6$C$_6$ monolayer, in which two Mn atoms in a Mn$_2$N$_6$ moiety are considered as a dimer and the distance between two dimers is short. Its structural stability is demonstrated by the phonon spectra and molecular dynamics calculations and the bonding characters are analyzed.
Then, we compute the critical temperature based on Heisenberg model with Monte Carlo simulation method, and the result manifest that Mn$_2$N$_6$C$_6$ monolayer is a near-room-temperature ferromagnetic 2D material with Curie temperatures of 272 K.
The example of Mn$_2$N$_6$C$_6$ monolayer is a powerful evidence to verify the feasibility of our proposed approach.

We sincerely thank Prof. Liang Liu in Shandong University for very helpful discussions on MC calculations. This work was supported by the National Natural Science Foundation of China (Grants Nos. 11974207, 11974194, 11774420, 12074040), the National R\&D
Program of China (Grants Nos. 2016YFA0300503, 2017YFA0302900), and the Major Basic Program of Natural Science Foundation of Shandong Province (Grant No. ZR2021ZD01).

\bibliography{Ref}

\begin{thebibliography}{48}%
\makeatletter
\providecommand \@ifxundefined [1]{%
 \@ifx{#1\undefined}
}%
\providecommand \@ifnum [1]{%
 \ifnum #1\expandafter \@firstoftwo
 \else \expandafter \@secondoftwo
 \fi
}%
\providecommand \@ifx [1]{%
 \ifx #1\expandafter \@firstoftwo
 \else \expandafter \@secondoftwo
 \fi
}%
\providecommand \natexlab [1]{#1}%
\providecommand \enquote  [1]{``#1''}%
\providecommand \bibnamefont  [1]{#1}%
\providecommand \bibfnamefont [1]{#1}%
\providecommand \citenamefont [1]{#1}%
\providecommand \href@noop [0]{\@secondoftwo}%
\providecommand \href [0]{\begingroup \@sanitize@url \@href}%
\providecommand \@href[1]{\@@startlink{#1}\@@href}%
\providecommand \@@href[1]{\endgroup#1\@@endlink}%
\providecommand \@sanitize@url [0]{\catcode `\\12\catcode `\$12\catcode
  `\&12\catcode `\#12\catcode `\^12\catcode `\_12\catcode `\%12\relax}%
\providecommand \@@startlink[1]{}%
\providecommand \@@endlink[0]{}%
\providecommand \url  [0]{\begingroup\@sanitize@url \@url }%
\providecommand \@url [1]{\endgroup\@href {#1}{\urlprefix }}%
\providecommand \urlprefix  [0]{URL }%
\providecommand \Eprint [0]{\href }%
\providecommand \doibase [0]{http://dx.doi.org/}%
\providecommand \selectlanguage [0]{\@gobble}%
\providecommand \bibinfo  [0]{\@secondoftwo}%
\providecommand \bibfield  [0]{\@secondoftwo}%
\providecommand \translation [1]{[#1]}%
\providecommand \BibitemOpen [0]{}%
\providecommand \bibitemStop [0]{}%
\providecommand \bibitemNoStop [0]{.\EOS\space}%
\providecommand \EOS [0]{\spacefactor3000\relax}%
\providecommand \BibitemShut  [1]{\csname bibitem#1\endcsname}%
\let\auto@bib@innerbib\@empty
\bibitem [{\citenamefont {Novoselov}(2004)}]{Novoselov2004}%
  \BibitemOpen
  \bibfield  {author} {\bibinfo {author} {\bibfnamefont {K.~S.}\ \bibnamefont
  {Novoselov}},\ }\href {\doibase 10.1126/science.1102896} {\bibfield
  {journal} {\bibinfo  {journal} {Science}\ }\textbf {\bibinfo {volume}
  {306}},\ \bibinfo {pages} {666} (\bibinfo {year} {2004})}\BibitemShut
  {NoStop}%
\bibitem [{\citenamefont {Song}\ \emph {et~al.}(2010)\citenamefont {Song},
  \citenamefont {Ci}, \citenamefont {Lu}, \citenamefont {Sorokin},
  \citenamefont {Jin}, \citenamefont {Ni}, \citenamefont {Kvashnin},
  \citenamefont {Kvashnin}, \citenamefont {Lou}, \citenamefont {Yakobson},\
  and\ \citenamefont {Ajayan}}]{Song2010}%
  \BibitemOpen
  \bibfield  {author} {\bibinfo {author} {\bibfnamefont {L.}~\bibnamefont
  {Song}}, \bibinfo {author} {\bibfnamefont {L.}~\bibnamefont {Ci}}, \bibinfo
  {author} {\bibfnamefont {H.}~\bibnamefont {Lu}}, \bibinfo {author}
  {\bibfnamefont {P.~B.}\ \bibnamefont {Sorokin}}, \bibinfo {author}
  {\bibfnamefont {C.}~\bibnamefont {Jin}}, \bibinfo {author} {\bibfnamefont
  {J.}~\bibnamefont {Ni}}, \bibinfo {author} {\bibfnamefont {A.~G.}\
  \bibnamefont {Kvashnin}}, \bibinfo {author} {\bibfnamefont {D.~G.}\
  \bibnamefont {Kvashnin}}, \bibinfo {author} {\bibfnamefont {J.}~\bibnamefont
  {Lou}}, \bibinfo {author} {\bibfnamefont {B.~I.}\ \bibnamefont {Yakobson}}, \
  and\ \bibinfo {author} {\bibfnamefont {P.~M.}\ \bibnamefont {Ajayan}},\
  }\href {\doibase 10.1021/nl1022139} {\bibfield  {journal} {\bibinfo
  {journal} {Nano Letters}\ }\textbf {\bibinfo {volume} {10}},\ \bibinfo
  {pages} {3209} (\bibinfo {year} {2010})}\BibitemShut {NoStop}%
\bibitem [{\citenamefont {Lalmi}\ \emph {et~al.}(2010)\citenamefont {Lalmi},
  \citenamefont {Oughaddou}, \citenamefont {Enriquez}, \citenamefont {Kara},
  \citenamefont {Vizzini}, \citenamefont {Ealet},\ and\ \citenamefont
  {Aufray}}]{Lalmi2010}%
  \BibitemOpen
  \bibfield  {author} {\bibinfo {author} {\bibfnamefont {B.}~\bibnamefont
  {Lalmi}}, \bibinfo {author} {\bibfnamefont {H.}~\bibnamefont {Oughaddou}},
  \bibinfo {author} {\bibfnamefont {H.}~\bibnamefont {Enriquez}}, \bibinfo
  {author} {\bibfnamefont {A.}~\bibnamefont {Kara}}, \bibinfo {author}
  {\bibfnamefont {S.}~\bibnamefont {Vizzini}}, \bibinfo {author} {\bibfnamefont
  {B.}~\bibnamefont {Ealet}}, \ and\ \bibinfo {author} {\bibfnamefont
  {B.}~\bibnamefont {Aufray}},\ }\href {\doibase 10.1063/1.3524215} {\bibfield
  {journal} {\bibinfo  {journal} {Applied Physics Letters}\ }\textbf {\bibinfo
  {volume} {97}},\ \bibinfo {pages} {223109} (\bibinfo {year}
  {2010})}\BibitemShut {NoStop}%
\bibitem [{\citenamefont {Mannix}\ \emph {et~al.}(2015)\citenamefont {Mannix},
  \citenamefont {Zhou}, \citenamefont {Kiraly}, \citenamefont {Wood},
  \citenamefont {Alducin}, \citenamefont {Myers}, \citenamefont {Liu},
  \citenamefont {Fisher}, \citenamefont {Santiago}, \citenamefont {Guest},
  \citenamefont {Yacaman}, \citenamefont {Ponce}, \citenamefont {Oganov},
  \citenamefont {Hersam},\ and\ \citenamefont {Guisinger}}]{Mannix2015}%
  \BibitemOpen
  \bibfield  {author} {\bibinfo {author} {\bibfnamefont {A.~J.}\ \bibnamefont
  {Mannix}}, \bibinfo {author} {\bibfnamefont {X.-F.}\ \bibnamefont {Zhou}},
  \bibinfo {author} {\bibfnamefont {B.}~\bibnamefont {Kiraly}}, \bibinfo
  {author} {\bibfnamefont {J.~D.}\ \bibnamefont {Wood}}, \bibinfo {author}
  {\bibfnamefont {D.}~\bibnamefont {Alducin}}, \bibinfo {author} {\bibfnamefont
  {B.~D.}\ \bibnamefont {Myers}}, \bibinfo {author} {\bibfnamefont
  {X.}~\bibnamefont {Liu}}, \bibinfo {author} {\bibfnamefont {B.~L.}\
  \bibnamefont {Fisher}}, \bibinfo {author} {\bibfnamefont {U.}~\bibnamefont
  {Santiago}}, \bibinfo {author} {\bibfnamefont {J.~R.}\ \bibnamefont {Guest}},
  \bibinfo {author} {\bibfnamefont {M.~J.}\ \bibnamefont {Yacaman}}, \bibinfo
  {author} {\bibfnamefont {A.}~\bibnamefont {Ponce}}, \bibinfo {author}
  {\bibfnamefont {A.~R.}\ \bibnamefont {Oganov}}, \bibinfo {author}
  {\bibfnamefont {M.~C.}\ \bibnamefont {Hersam}}, \ and\ \bibinfo {author}
  {\bibfnamefont {N.~P.}\ \bibnamefont {Guisinger}},\ }\href {\doibase
  10.1126/science.aad1080} {\bibfield  {journal} {\bibinfo  {journal}
  {Science}\ }\textbf {\bibinfo {volume} {350}},\ \bibinfo {pages} {1513}
  (\bibinfo {year} {2015})}\BibitemShut {NoStop}%
\bibitem [{\citenamefont {Saxena}\ \emph {et~al.}(2016)\citenamefont {Saxena},
  \citenamefont {Chaudhary},\ and\ \citenamefont {Shukla}}]{Saxena2016}%
  \BibitemOpen
  \bibfield  {author} {\bibinfo {author} {\bibfnamefont {S.}~\bibnamefont
  {Saxena}}, \bibinfo {author} {\bibfnamefont {R.~P.}\ \bibnamefont
  {Chaudhary}}, \ and\ \bibinfo {author} {\bibfnamefont {S.}~\bibnamefont
  {Shukla}},\ }\href {\doibase 10.1038/srep31073} {\bibfield  {journal}
  {\bibinfo  {journal} {Scientific Reports}\ }\textbf {\bibinfo {volume} {6}},\
  \bibinfo {pages} {31073} (\bibinfo {year} {2016})}\BibitemShut {NoStop}%
\bibitem [{\citenamefont {Coleman}\ \emph {et~al.}(2011)\citenamefont
  {Coleman}, \citenamefont {Lotya}, \citenamefont {O'Neill}, \citenamefont
  {Bergin}, \citenamefont {King}, \citenamefont {Khan}, \citenamefont {Young},
  \citenamefont {Gaucher}, \citenamefont {De}, \citenamefont {Smith},
  \citenamefont {Shvets}, \citenamefont {Arora}, \citenamefont {Stanton},
  \citenamefont {Kim}, \citenamefont {Lee}, \citenamefont {Kim}, \citenamefont
  {Duesberg}, \citenamefont {Hallam}, \citenamefont {Boland}, \citenamefont
  {Wang}, \citenamefont {Donegan}, \citenamefont {Grunlan}, \citenamefont
  {Moriarty}, \citenamefont {Shmeliov}, \citenamefont {Nicholls}, \citenamefont
  {Perkins}, \citenamefont {Grieveson}, \citenamefont {Theuwissen},
  \citenamefont {McComb}, \citenamefont {Nellist},\ and\ \citenamefont
  {Nicolosi}}]{Coleman2011}%
  \BibitemOpen
  \bibfield  {author} {\bibinfo {author} {\bibfnamefont {J.~N.}\ \bibnamefont
  {Coleman}}, \bibinfo {author} {\bibfnamefont {M.}~\bibnamefont {Lotya}},
  \bibinfo {author} {\bibfnamefont {A.}~\bibnamefont {O'Neill}}, \bibinfo
  {author} {\bibfnamefont {S.~D.}\ \bibnamefont {Bergin}}, \bibinfo {author}
  {\bibfnamefont {P.~J.}\ \bibnamefont {King}}, \bibinfo {author}
  {\bibfnamefont {U.}~\bibnamefont {Khan}}, \bibinfo {author} {\bibfnamefont
  {K.}~\bibnamefont {Young}}, \bibinfo {author} {\bibfnamefont
  {A.}~\bibnamefont {Gaucher}}, \bibinfo {author} {\bibfnamefont
  {S.}~\bibnamefont {De}}, \bibinfo {author} {\bibfnamefont {R.~J.}\
  \bibnamefont {Smith}}, \bibinfo {author} {\bibfnamefont {I.~V.}\ \bibnamefont
  {Shvets}}, \bibinfo {author} {\bibfnamefont {S.~K.}\ \bibnamefont {Arora}},
  \bibinfo {author} {\bibfnamefont {G.}~\bibnamefont {Stanton}}, \bibinfo
  {author} {\bibfnamefont {H.-Y.}\ \bibnamefont {Kim}}, \bibinfo {author}
  {\bibfnamefont {K.}~\bibnamefont {Lee}}, \bibinfo {author} {\bibfnamefont
  {G.~T.}\ \bibnamefont {Kim}}, \bibinfo {author} {\bibfnamefont {G.~S.}\
  \bibnamefont {Duesberg}}, \bibinfo {author} {\bibfnamefont {T.}~\bibnamefont
  {Hallam}}, \bibinfo {author} {\bibfnamefont {J.~J.}\ \bibnamefont {Boland}},
  \bibinfo {author} {\bibfnamefont {J.~J.}\ \bibnamefont {Wang}}, \bibinfo
  {author} {\bibfnamefont {J.~F.}\ \bibnamefont {Donegan}}, \bibinfo {author}
  {\bibfnamefont {J.~C.}\ \bibnamefont {Grunlan}}, \bibinfo {author}
  {\bibfnamefont {G.}~\bibnamefont {Moriarty}}, \bibinfo {author}
  {\bibfnamefont {A.}~\bibnamefont {Shmeliov}}, \bibinfo {author}
  {\bibfnamefont {R.~J.}\ \bibnamefont {Nicholls}}, \bibinfo {author}
  {\bibfnamefont {J.~M.}\ \bibnamefont {Perkins}}, \bibinfo {author}
  {\bibfnamefont {E.~M.}\ \bibnamefont {Grieveson}}, \bibinfo {author}
  {\bibfnamefont {K.}~\bibnamefont {Theuwissen}}, \bibinfo {author}
  {\bibfnamefont {D.~W.}\ \bibnamefont {McComb}}, \bibinfo {author}
  {\bibfnamefont {P.~D.}\ \bibnamefont {Nellist}}, \ and\ \bibinfo {author}
  {\bibfnamefont {V.}~\bibnamefont {Nicolosi}},\ }\href {\doibase
  10.1126/science.1194975} {\bibfield  {journal} {\bibinfo  {journal}
  {Science}\ }\textbf {\bibinfo {volume} {331}},\ \bibinfo {pages} {568}
  (\bibinfo {year} {2011})}\BibitemShut {NoStop}%
\bibitem [{\citenamefont {Naguib}\ \emph {et~al.}(2011)\citenamefont {Naguib},
  \citenamefont {Kurtoglu}, \citenamefont {Presser}, \citenamefont {Lu},
  \citenamefont {Niu}, \citenamefont {Heon}, \citenamefont {Hultman},
  \citenamefont {Gogotsi},\ and\ \citenamefont {Barsoum}}]{Naguib2011}%
  \BibitemOpen
  \bibfield  {author} {\bibinfo {author} {\bibfnamefont {M.}~\bibnamefont
  {Naguib}}, \bibinfo {author} {\bibfnamefont {M.}~\bibnamefont {Kurtoglu}},
  \bibinfo {author} {\bibfnamefont {V.}~\bibnamefont {Presser}}, \bibinfo
  {author} {\bibfnamefont {J.}~\bibnamefont {Lu}}, \bibinfo {author}
  {\bibfnamefont {J.}~\bibnamefont {Niu}}, \bibinfo {author} {\bibfnamefont
  {M.}~\bibnamefont {Heon}}, \bibinfo {author} {\bibfnamefont {L.}~\bibnamefont
  {Hultman}}, \bibinfo {author} {\bibfnamefont {Y.}~\bibnamefont {Gogotsi}}, \
  and\ \bibinfo {author} {\bibfnamefont {M.~W.}\ \bibnamefont {Barsoum}},\
  }\href {\doibase 10.1002/adma.201102306} {\bibfield  {journal} {\bibinfo
  {journal} {Advanced Materials}\ }\textbf {\bibinfo {volume} {23}},\ \bibinfo
  {pages} {4248} (\bibinfo {year} {2011})}\BibitemShut {NoStop}%
\bibitem [{\citenamefont {Andriotis}\ \emph {et~al.}(2016)\citenamefont
  {Andriotis}, \citenamefont {Richter},\ and\ \citenamefont
  {Menon}}]{Andriotis2016}%
  \BibitemOpen
  \bibfield  {author} {\bibinfo {author} {\bibfnamefont {A.~N.}\ \bibnamefont
  {Andriotis}}, \bibinfo {author} {\bibfnamefont {E.}~\bibnamefont {Richter}},
  \ and\ \bibinfo {author} {\bibfnamefont {M.}~\bibnamefont {Menon}},\ }\href
  {\doibase 10.1103/PhysRevB.93.081413} {\bibfield  {journal} {\bibinfo
  {journal} {Physical Review B}\ }\textbf {\bibinfo {volume} {93}},\ \bibinfo
  {pages} {081413(R)} (\bibinfo {year} {2016})}\BibitemShut {NoStop}%
\bibitem [{\citenamefont {Wang}\ \emph {et~al.}(2017)\citenamefont {Wang},
  \citenamefont {Yuan}, \citenamefont {Li}, \citenamefont {Shi},\ and\
  \citenamefont {Wang}}]{Wang2017}%
  \BibitemOpen
  \bibfield  {author} {\bibinfo {author} {\bibfnamefont {B.}~\bibnamefont
  {Wang}}, \bibinfo {author} {\bibfnamefont {S.}~\bibnamefont {Yuan}}, \bibinfo
  {author} {\bibfnamefont {Y.}~\bibnamefont {Li}}, \bibinfo {author}
  {\bibfnamefont {L.}~\bibnamefont {Shi}}, \ and\ \bibinfo {author}
  {\bibfnamefont {J.}~\bibnamefont {Wang}},\ }\href {\doibase
  10.1039/C7NR00455A} {\bibfield  {journal} {\bibinfo  {journal} {Nanoscale}\
  }\textbf {\bibinfo {volume} {9}},\ \bibinfo {pages} {5577} (\bibinfo {year}
  {2017})}\BibitemShut {NoStop}%
\bibitem [{\citenamefont {Sun}\ \emph {et~al.}(2017)\citenamefont {Sun},
  \citenamefont {Zhuo}, \citenamefont {Wu},\ and\ \citenamefont
  {Yang}}]{Sun2017}%
  \BibitemOpen
  \bibfield  {author} {\bibinfo {author} {\bibfnamefont {Y.}~\bibnamefont
  {Sun}}, \bibinfo {author} {\bibfnamefont {Z.}~\bibnamefont {Zhuo}}, \bibinfo
  {author} {\bibfnamefont {X.}~\bibnamefont {Wu}}, \ and\ \bibinfo {author}
  {\bibfnamefont {J.}~\bibnamefont {Yang}},\ }\href {\doibase
  10.1021/acs.nanolett.6b04884} {\bibfield  {journal} {\bibinfo  {journal}
  {Nano Letters}\ }\textbf {\bibinfo {volume} {17}},\ \bibinfo {pages} {2771}
  (\bibinfo {year} {2017})}\BibitemShut {NoStop}%
\bibitem [{\citenamefont {Tang}\ \emph {et~al.}(2019)\citenamefont {Tang},
  \citenamefont {Sun}, \citenamefont {Gu}, \citenamefont {Lu}, \citenamefont
  {Kou},\ and\ \citenamefont {Chen}}]{Tang2019}%
  \BibitemOpen
  \bibfield  {author} {\bibinfo {author} {\bibfnamefont {X.}~\bibnamefont
  {Tang}}, \bibinfo {author} {\bibfnamefont {W.}~\bibnamefont {Sun}}, \bibinfo
  {author} {\bibfnamefont {Y.}~\bibnamefont {Gu}}, \bibinfo {author}
  {\bibfnamefont {C.}~\bibnamefont {Lu}}, \bibinfo {author} {\bibfnamefont
  {L.}~\bibnamefont {Kou}}, \ and\ \bibinfo {author} {\bibfnamefont
  {C.}~\bibnamefont {Chen}},\ }\href {\doibase 10.1103/PhysRevB.99.045445}
  {\bibfield  {journal} {\bibinfo  {journal} {Physical Review B}\ }\textbf
  {\bibinfo {volume} {99}},\ \bibinfo {pages} {045445} (\bibinfo {year}
  {2019})}\BibitemShut {NoStop}%
\bibitem [{\citenamefont {Huang}\ \emph {et~al.}(2017)\citenamefont {Huang},
  \citenamefont {Clark}, \citenamefont {Navarro-Moratalla}, \citenamefont
  {Klein}, \citenamefont {Cheng}, \citenamefont {Seyler}, \citenamefont
  {Zhong}, \citenamefont {Schmidgall}, \citenamefont {McGuire}, \citenamefont
  {Cobden}, \citenamefont {Yao}, \citenamefont {Xiao}, \citenamefont
  {Jarillo-Herrero},\ and\ \citenamefont {Xu}}]{Huang2017}%
  \BibitemOpen
  \bibfield  {author} {\bibinfo {author} {\bibfnamefont {B.}~\bibnamefont
  {Huang}}, \bibinfo {author} {\bibfnamefont {G.}~\bibnamefont {Clark}},
  \bibinfo {author} {\bibfnamefont {E.}~\bibnamefont {Navarro-Moratalla}},
  \bibinfo {author} {\bibfnamefont {D.~R.}\ \bibnamefont {Klein}}, \bibinfo
  {author} {\bibfnamefont {R.}~\bibnamefont {Cheng}}, \bibinfo {author}
  {\bibfnamefont {K.~L.}\ \bibnamefont {Seyler}}, \bibinfo {author}
  {\bibfnamefont {D.}~\bibnamefont {Zhong}}, \bibinfo {author} {\bibfnamefont
  {E.}~\bibnamefont {Schmidgall}}, \bibinfo {author} {\bibfnamefont {M.~A.}\
  \bibnamefont {McGuire}}, \bibinfo {author} {\bibfnamefont {D.~H.}\
  \bibnamefont {Cobden}}, \bibinfo {author} {\bibfnamefont {W.}~\bibnamefont
  {Yao}}, \bibinfo {author} {\bibfnamefont {D.}~\bibnamefont {Xiao}}, \bibinfo
  {author} {\bibfnamefont {P.}~\bibnamefont {Jarillo-Herrero}}, \ and\ \bibinfo
  {author} {\bibfnamefont {X.}~\bibnamefont {Xu}},\ }\href {\doibase
  10.1038/nature22391} {\bibfield  {journal} {\bibinfo  {journal} {Nature}\
  }\textbf {\bibinfo {volume} {546}},\ \bibinfo {pages} {270} (\bibinfo {year}
  {2017})}\BibitemShut {NoStop}%
\bibitem [{\citenamefont {Gong}\ \emph {et~al.}(2017)\citenamefont {Gong},
  \citenamefont {Li}, \citenamefont {Li}, \citenamefont {Ji}, \citenamefont
  {Stern}, \citenamefont {Xia}, \citenamefont {Cao}, \citenamefont {Bao},
  \citenamefont {Wang}, \citenamefont {Wang}, \citenamefont {Qiu},
  \citenamefont {Cava}, \citenamefont {Louie}, \citenamefont {Xia},\ and\
  \citenamefont {Zhang}}]{Gong2017}%
  \BibitemOpen
  \bibfield  {author} {\bibinfo {author} {\bibfnamefont {C.}~\bibnamefont
  {Gong}}, \bibinfo {author} {\bibfnamefont {L.}~\bibnamefont {Li}}, \bibinfo
  {author} {\bibfnamefont {Z.}~\bibnamefont {Li}}, \bibinfo {author}
  {\bibfnamefont {H.}~\bibnamefont {Ji}}, \bibinfo {author} {\bibfnamefont
  {A.}~\bibnamefont {Stern}}, \bibinfo {author} {\bibfnamefont
  {Y.}~\bibnamefont {Xia}}, \bibinfo {author} {\bibfnamefont {T.}~\bibnamefont
  {Cao}}, \bibinfo {author} {\bibfnamefont {W.}~\bibnamefont {Bao}}, \bibinfo
  {author} {\bibfnamefont {C.}~\bibnamefont {Wang}}, \bibinfo {author}
  {\bibfnamefont {Y.}~\bibnamefont {Wang}}, \bibinfo {author} {\bibfnamefont
  {Z.~Q.}\ \bibnamefont {Qiu}}, \bibinfo {author} {\bibfnamefont {R.~J.}\
  \bibnamefont {Cava}}, \bibinfo {author} {\bibfnamefont {S.~G.}\ \bibnamefont
  {Louie}}, \bibinfo {author} {\bibfnamefont {J.}~\bibnamefont {Xia}}, \ and\
  \bibinfo {author} {\bibfnamefont {X.}~\bibnamefont {Zhang}},\ }\href
  {\doibase 10.1038/nature22060} {\bibfield  {journal} {\bibinfo  {journal}
  {Nature}\ }\textbf {\bibinfo {volume} {546}},\ \bibinfo {pages} {265}
  (\bibinfo {year} {2017})}\BibitemShut {NoStop}%
\bibitem [{\citenamefont {Bonilla}\ \emph {et~al.}(2018)\citenamefont
  {Bonilla}, \citenamefont {Kolekar}, \citenamefont {Ma}, \citenamefont {Diaz},
  \citenamefont {Kalappattil}, \citenamefont {Das}, \citenamefont {Eggers},
  \citenamefont {Gutierrez}, \citenamefont {Phan},\ and\ \citenamefont
  {Batzill}}]{Bonilla2018}%
  \BibitemOpen
  \bibfield  {author} {\bibinfo {author} {\bibfnamefont {M.}~\bibnamefont
  {Bonilla}}, \bibinfo {author} {\bibfnamefont {S.}~\bibnamefont {Kolekar}},
  \bibinfo {author} {\bibfnamefont {Y.}~\bibnamefont {Ma}}, \bibinfo {author}
  {\bibfnamefont {H.~C.}\ \bibnamefont {Diaz}}, \bibinfo {author}
  {\bibfnamefont {V.}~\bibnamefont {Kalappattil}}, \bibinfo {author}
  {\bibfnamefont {R.}~\bibnamefont {Das}}, \bibinfo {author} {\bibfnamefont
  {T.}~\bibnamefont {Eggers}}, \bibinfo {author} {\bibfnamefont {H.~R.}\
  \bibnamefont {Gutierrez}}, \bibinfo {author} {\bibfnamefont {M.-h.}\
  \bibnamefont {Phan}}, \ and\ \bibinfo {author} {\bibfnamefont
  {M.}~\bibnamefont {Batzill}},\ }\href {\doibase 10.1038/s41565-018-0063-9}
  {\bibfield  {journal} {\bibinfo  {journal} {Nature Nanotechnology}\ }\textbf
  {\bibinfo {volume} {13}},\ \bibinfo {pages} {289} (\bibinfo {year}
  {2018})}\BibitemShut {NoStop}%
\bibitem [{\citenamefont {Hu}\ \emph {et~al.}(2021)\citenamefont {Hu},
  \citenamefont {Wang}, \citenamefont {Tan}, \citenamefont {Duan},
  \citenamefont {Li}, \citenamefont {Li}, \citenamefont {Ji}, \citenamefont
  {Lu}, \citenamefont {Wang}, \citenamefont {Sun}, \citenamefont {Hu},\ and\
  \citenamefont {Yan}}]{Hu2021}%
  \BibitemOpen
  \bibfield  {author} {\bibinfo {author} {\bibfnamefont {W.}~\bibnamefont
  {Hu}}, \bibinfo {author} {\bibfnamefont {C.}~\bibnamefont {Wang}}, \bibinfo
  {author} {\bibfnamefont {H.}~\bibnamefont {Tan}}, \bibinfo {author}
  {\bibfnamefont {H.}~\bibnamefont {Duan}}, \bibinfo {author} {\bibfnamefont
  {G.}~\bibnamefont {Li}}, \bibinfo {author} {\bibfnamefont {N.}~\bibnamefont
  {Li}}, \bibinfo {author} {\bibfnamefont {Q.}~\bibnamefont {Ji}}, \bibinfo
  {author} {\bibfnamefont {Y.}~\bibnamefont {Lu}}, \bibinfo {author}
  {\bibfnamefont {Y.}~\bibnamefont {Wang}}, \bibinfo {author} {\bibfnamefont
  {Z.}~\bibnamefont {Sun}}, \bibinfo {author} {\bibfnamefont {F.}~\bibnamefont
  {Hu}}, \ and\ \bibinfo {author} {\bibfnamefont {W.}~\bibnamefont {Yan}},\
  }\href {\doibase 10.1038/s41467-021-22122-2} {\bibfield  {journal} {\bibinfo
  {journal} {Nature Communications}\ }\textbf {\bibinfo {volume} {12}},\
  \bibinfo {pages} {1854} (\bibinfo {year} {2021})}\BibitemShut {NoStop}%
\bibitem [{\citenamefont {Zhang}\ \emph {et~al.}(2015)\citenamefont {Zhang},
  \citenamefont {Qu}, \citenamefont {Zhu},\ and\ \citenamefont
  {Lam}}]{Zhang2015}%
  \BibitemOpen
  \bibfield  {author} {\bibinfo {author} {\bibfnamefont {W.-B.}\ \bibnamefont
  {Zhang}}, \bibinfo {author} {\bibfnamefont {Q.}~\bibnamefont {Qu}}, \bibinfo
  {author} {\bibfnamefont {P.}~\bibnamefont {Zhu}}, \ and\ \bibinfo {author}
  {\bibfnamefont {C.-H.}\ \bibnamefont {Lam}},\ }\href {\doibase
  10.1039/C5TC02840J} {\bibfield  {journal} {\bibinfo  {journal} {Journal of
  Materials Chemistry C}\ }\textbf {\bibinfo {volume} {3}},\ \bibinfo {pages}
  {12457} (\bibinfo {year} {2015})}\BibitemShut {NoStop}%
\bibitem [{\citenamefont {Sivadas}\ \emph {et~al.}(2015)\citenamefont
  {Sivadas}, \citenamefont {Daniels}, \citenamefont {Swendsen}, \citenamefont
  {Okamoto},\ and\ \citenamefont {Xiao}}]{Sivadas2015}%
  \BibitemOpen
  \bibfield  {author} {\bibinfo {author} {\bibfnamefont {N.}~\bibnamefont
  {Sivadas}}, \bibinfo {author} {\bibfnamefont {M.~W.}\ \bibnamefont
  {Daniels}}, \bibinfo {author} {\bibfnamefont {R.~H.}\ \bibnamefont
  {Swendsen}}, \bibinfo {author} {\bibfnamefont {S.}~\bibnamefont {Okamoto}}, \
  and\ \bibinfo {author} {\bibfnamefont {D.}~\bibnamefont {Xiao}},\ }\href
  {\doibase 10.1103/PhysRevB.91.235425} {\bibfield  {journal} {\bibinfo
  {journal} {Physical Review B}\ }\textbf {\bibinfo {volume} {91}},\ \bibinfo
  {pages} {235425} (\bibinfo {year} {2015})}\BibitemShut {NoStop}%
\bibitem [{\citenamefont {Guan}\ \emph {et~al.}(2019)\citenamefont {Guan},
  \citenamefont {Huang}, \citenamefont {Deng},\ and\ \citenamefont
  {Kan}}]{Guan2019}%
  \BibitemOpen
  \bibfield  {author} {\bibinfo {author} {\bibfnamefont {J.}~\bibnamefont
  {Guan}}, \bibinfo {author} {\bibfnamefont {C.}~\bibnamefont {Huang}},
  \bibinfo {author} {\bibfnamefont {K.}~\bibnamefont {Deng}}, \ and\ \bibinfo
  {author} {\bibfnamefont {E.}~\bibnamefont {Kan}},\ }\href {\doibase
  10.1021/acs.jpcc.9b00763} {\bibfield  {journal} {\bibinfo  {journal} {The
  Journal of Physical Chemistry C}\ }\textbf {\bibinfo {volume} {123}},\
  \bibinfo {pages} {10114} (\bibinfo {year} {2019})}\BibitemShut {NoStop}%
\bibitem [{\citenamefont {Zheng}\ \emph {et~al.}(2019)\citenamefont {Zheng},
  \citenamefont {Huang}, \citenamefont {Yu}, \citenamefont {Xu}, \citenamefont
  {Zhang}, \citenamefont {Xu}, \citenamefont {Liu}, \citenamefont {Kan},
  \citenamefont {Wang},\ and\ \citenamefont {Yang}}]{Zheng2019}%
  \BibitemOpen
  \bibfield  {author} {\bibinfo {author} {\bibfnamefont {S.}~\bibnamefont
  {Zheng}}, \bibinfo {author} {\bibfnamefont {C.}~\bibnamefont {Huang}},
  \bibinfo {author} {\bibfnamefont {T.}~\bibnamefont {Yu}}, \bibinfo {author}
  {\bibfnamefont {M.}~\bibnamefont {Xu}}, \bibinfo {author} {\bibfnamefont
  {S.}~\bibnamefont {Zhang}}, \bibinfo {author} {\bibfnamefont
  {H.}~\bibnamefont {Xu}}, \bibinfo {author} {\bibfnamefont {Y.}~\bibnamefont
  {Liu}}, \bibinfo {author} {\bibfnamefont {E.}~\bibnamefont {Kan}}, \bibinfo
  {author} {\bibfnamefont {Y.}~\bibnamefont {Wang}}, \ and\ \bibinfo {author}
  {\bibfnamefont {G.}~\bibnamefont {Yang}},\ }\href {\doibase
  10.1021/acs.jpclett.9b00970} {\bibfield  {journal} {\bibinfo  {journal} {The
  Journal of Physical Chemistry Letters}\ }\textbf {\bibinfo {volume} {10}},\
  \bibinfo {pages} {2733} (\bibinfo {year} {2019})}\BibitemShut {NoStop}%
\bibitem [{\citenamefont {Chen}\ \emph {et~al.}(2020)\citenamefont {Chen},
  \citenamefont {Wu}, \citenamefont {Li}, \citenamefont {Sun}, \citenamefont
  {Ding}, \citenamefont {Huang},\ and\ \citenamefont {Kan}}]{Chen2020}%
  \BibitemOpen
  \bibfield  {author} {\bibinfo {author} {\bibfnamefont {S.}~\bibnamefont
  {Chen}}, \bibinfo {author} {\bibfnamefont {F.}~\bibnamefont {Wu}}, \bibinfo
  {author} {\bibfnamefont {Q.}~\bibnamefont {Li}}, \bibinfo {author}
  {\bibfnamefont {H.}~\bibnamefont {Sun}}, \bibinfo {author} {\bibfnamefont
  {J.}~\bibnamefont {Ding}}, \bibinfo {author} {\bibfnamefont {C.}~\bibnamefont
  {Huang}}, \ and\ \bibinfo {author} {\bibfnamefont {E.}~\bibnamefont {Kan}},\
  }\href {\doibase 10.1039/D0NR03340E} {\bibfield  {journal} {\bibinfo
  {journal} {Nanoscale}\ }\textbf {\bibinfo {volume} {12}},\ \bibinfo {pages}
  {15670} (\bibinfo {year} {2020})}\BibitemShut {NoStop}%
\bibitem [{\citenamefont {Wang}\ \emph {et~al.}(2020)\citenamefont {Wang},
  \citenamefont {Zhang}, \citenamefont {Zhang}, \citenamefont {Yuan},
  \citenamefont {Guo}, \citenamefont {Dong},\ and\ \citenamefont
  {Wang}}]{Wang2020}%
  \BibitemOpen
  \bibfield  {author} {\bibinfo {author} {\bibfnamefont {B.}~\bibnamefont
  {Wang}}, \bibinfo {author} {\bibfnamefont {X.}~\bibnamefont {Zhang}},
  \bibinfo {author} {\bibfnamefont {Y.}~\bibnamefont {Zhang}}, \bibinfo
  {author} {\bibfnamefont {S.}~\bibnamefont {Yuan}}, \bibinfo {author}
  {\bibfnamefont {Y.}~\bibnamefont {Guo}}, \bibinfo {author} {\bibfnamefont
  {S.}~\bibnamefont {Dong}}, \ and\ \bibinfo {author} {\bibfnamefont
  {J.}~\bibnamefont {Wang}},\ }\href {\doibase 10.1039/D0MH00183J} {\bibfield
  {journal} {\bibinfo  {journal} {Materials Horizons}\ }\textbf {\bibinfo
  {volume} {7}},\ \bibinfo {pages} {1623} (\bibinfo {year} {2020})}\BibitemShut
  {NoStop}%
\bibitem [{\citenamefont {Liu}\ \emph {et~al.}(2021{\natexlab{a}})\citenamefont
  {Liu}, \citenamefont {Zhang}, \citenamefont {Gao}, \citenamefont {Yan},\ and\
  \citenamefont {Xie}}]{Liu2021b}%
  \BibitemOpen
  \bibfield  {author} {\bibinfo {author} {\bibfnamefont {D.}~\bibnamefont
  {Liu}}, \bibinfo {author} {\bibfnamefont {S.}~\bibnamefont {Zhang}}, \bibinfo
  {author} {\bibfnamefont {M.}~\bibnamefont {Gao}}, \bibinfo {author}
  {\bibfnamefont {X.-W.}\ \bibnamefont {Yan}}, \ and\ \bibinfo {author}
  {\bibfnamefont {Z.~Y.}\ \bibnamefont {Xie}},\ }\href {\doibase
  10.1063/5.0054730} {\bibfield  {journal} {\bibinfo  {journal} {Applied
  Physics Letters}\ }\textbf {\bibinfo {volume} {118}},\ \bibinfo {pages}
  {223104} (\bibinfo {year} {2021}{\natexlab{a}})}\BibitemShut {NoStop}%
\bibitem [{\citenamefont {Guo}\ \emph {et~al.}(2020)\citenamefont {Guo},
  \citenamefont {Wang}, \citenamefont {Zhang}, \citenamefont {Yuan},
  \citenamefont {Ma},\ and\ \citenamefont {Wang}}]{Guo2020}%
  \BibitemOpen
  \bibfield  {author} {\bibinfo {author} {\bibfnamefont {Y.}~\bibnamefont
  {Guo}}, \bibinfo {author} {\bibfnamefont {B.}~\bibnamefont {Wang}}, \bibinfo
  {author} {\bibfnamefont {X.}~\bibnamefont {Zhang}}, \bibinfo {author}
  {\bibfnamefont {S.}~\bibnamefont {Yuan}}, \bibinfo {author} {\bibfnamefont
  {L.}~\bibnamefont {Ma}}, \ and\ \bibinfo {author} {\bibfnamefont
  {J.}~\bibnamefont {Wang}},\ }\href {\doibase 10.1002/inf2.12096} {\bibfield
  {journal} {\bibinfo  {journal} {InfoMat}\ }\textbf {\bibinfo {volume} {2}},\
  \bibinfo {pages} {639} (\bibinfo {year} {2020})}\BibitemShut {NoStop}%
\bibitem [{\citenamefont {Huang}\ \emph {et~al.}(2020)\citenamefont {Huang},
  \citenamefont {Zhang}, \citenamefont {Xu}, \citenamefont {Wang},
  \citenamefont {Zhang},\ and\ \citenamefont {Zhang}}]{Huang2020}%
  \BibitemOpen
  \bibfield  {author} {\bibinfo {author} {\bibfnamefont {P.}~\bibnamefont
  {Huang}}, \bibinfo {author} {\bibfnamefont {P.}~\bibnamefont {Zhang}},
  \bibinfo {author} {\bibfnamefont {S.}~\bibnamefont {Xu}}, \bibinfo {author}
  {\bibfnamefont {H.}~\bibnamefont {Wang}}, \bibinfo {author} {\bibfnamefont
  {X.}~\bibnamefont {Zhang}}, \ and\ \bibinfo {author} {\bibfnamefont
  {H.}~\bibnamefont {Zhang}},\ }\href {\doibase 10.1039/C9NR08890C} {\bibfield
  {journal} {\bibinfo  {journal} {Nanoscale}\ }\textbf {\bibinfo {volume}
  {12}},\ \bibinfo {pages} {2309} (\bibinfo {year} {2020})}\BibitemShut
  {NoStop}%
\bibitem [{\citenamefont {Lu}\ \emph {et~al.}(2019)\citenamefont {Lu},
  \citenamefont {Wang}, \citenamefont {Hu}, \citenamefont {Liu}, \citenamefont
  {Zhao}, \citenamefont {Yang}, \citenamefont {Li}, \citenamefont {Luo},
  \citenamefont {Chi}, \citenamefont {Jiang}, \citenamefont {Li}, \citenamefont
  {Mu}, \citenamefont {Liao}, \citenamefont {Zhang},\ and\ \citenamefont
  {Sun}}]{Lu2019}%
  \BibitemOpen
  \bibfield  {author} {\bibinfo {author} {\bibfnamefont {Z.}~\bibnamefont
  {Lu}}, \bibinfo {author} {\bibfnamefont {B.}~\bibnamefont {Wang}}, \bibinfo
  {author} {\bibfnamefont {Y.}~\bibnamefont {Hu}}, \bibinfo {author}
  {\bibfnamefont {W.}~\bibnamefont {Liu}}, \bibinfo {author} {\bibfnamefont
  {Y.}~\bibnamefont {Zhao}}, \bibinfo {author} {\bibfnamefont {R.}~\bibnamefont
  {Yang}}, \bibinfo {author} {\bibfnamefont {Z.}~\bibnamefont {Li}}, \bibinfo
  {author} {\bibfnamefont {J.}~\bibnamefont {Luo}}, \bibinfo {author}
  {\bibfnamefont {B.}~\bibnamefont {Chi}}, \bibinfo {author} {\bibfnamefont
  {Z.}~\bibnamefont {Jiang}}, \bibinfo {author} {\bibfnamefont
  {M.}~\bibnamefont {Li}}, \bibinfo {author} {\bibfnamefont {S.}~\bibnamefont
  {Mu}}, \bibinfo {author} {\bibfnamefont {S.}~\bibnamefont {Liao}}, \bibinfo
  {author} {\bibfnamefont {J.}~\bibnamefont {Zhang}}, \ and\ \bibinfo {author}
  {\bibfnamefont {X.}~\bibnamefont {Sun}},\ }\href {\doibase
  10.1002/ange.201810175} {\bibfield  {journal} {\bibinfo  {journal}
  {Angewandte Chemie}\ }\textbf {\bibinfo {volume} {131}},\ \bibinfo {pages}
  {2648} (\bibinfo {year} {2019})}\BibitemShut {NoStop}%
\bibitem [{\citenamefont {Yuan}\ \emph {et~al.}(2020)\citenamefont {Yuan},
  \citenamefont {Guo}, \citenamefont {Shi}, \citenamefont {Zhao}, \citenamefont
  {Jia}, \citenamefont {Wang}, \citenamefont {Sun}, \citenamefont {Wu},\ and\
  \citenamefont {Li}}]{Yuan2020}%
  \BibitemOpen
  \bibfield  {author} {\bibinfo {author} {\bibfnamefont {Q.~Q.}\ \bibnamefont
  {Yuan}}, \bibinfo {author} {\bibfnamefont {Z.}~\bibnamefont {Guo}}, \bibinfo
  {author} {\bibfnamefont {Z.~Q.}\ \bibnamefont {Shi}}, \bibinfo {author}
  {\bibfnamefont {H.}~\bibnamefont {Zhao}}, \bibinfo {author} {\bibfnamefont
  {Z.~Y.}\ \bibnamefont {Jia}}, \bibinfo {author} {\bibfnamefont
  {Q.}~\bibnamefont {Wang}}, \bibinfo {author} {\bibfnamefont {J.}~\bibnamefont
  {Sun}}, \bibinfo {author} {\bibfnamefont {D.}~\bibnamefont {Wu}}, \ and\
  \bibinfo {author} {\bibfnamefont {S.~C.}\ \bibnamefont {Li}},\ }\href
  {\doibase 10.1088/0256-307X/37/7/077502} {\bibfield  {journal} {\bibinfo
  {journal} {Chinese Physics Letters}\ }\textbf {\bibinfo {volume} {37}},\
  \bibinfo {pages} {077502} (\bibinfo {year} {2020})},\ \Eprint
  {http://arxiv.org/abs/2006.00333} {arXiv:2006.00333} \BibitemShut {NoStop}%
\bibitem [{\citenamefont {Kresse}\ and\ \citenamefont
  {Hafner}(1993)}]{PhysRevB.47.558}%
  \BibitemOpen
  \bibfield  {author} {\bibinfo {author} {\bibfnamefont {G.}~\bibnamefont
  {Kresse}}\ and\ \bibinfo {author} {\bibfnamefont {J.}~\bibnamefont
  {Hafner}},\ }\href {\doibase 10.1103/PhysRevB.47.558} {\bibfield  {journal}
  {\bibinfo  {journal} {Physical Review B}\ }\textbf {\bibinfo {volume} {47}},\
  \bibinfo {pages} {558} (\bibinfo {year} {1993})}\BibitemShut {NoStop}%
\bibitem [{\citenamefont {Kresse}\ and\ \citenamefont
  {Furthm{\"{u}}ller}(1996)}]{PhysRevB.54.11169}%
  \BibitemOpen
  \bibfield  {author} {\bibinfo {author} {\bibfnamefont {G.}~\bibnamefont
  {Kresse}}\ and\ \bibinfo {author} {\bibfnamefont {J.}~\bibnamefont
  {Furthm{\"{u}}ller}},\ }\href {\doibase 10.1103/PhysRevB.54.11169} {\bibfield
   {journal} {\bibinfo  {journal} {Phys. Rev. B}\ }\textbf {\bibinfo {volume}
  {54}},\ \bibinfo {pages} {11169} (\bibinfo {year} {1996})}\BibitemShut
  {NoStop}%
\bibitem [{\citenamefont {Perdew}\ \emph {et~al.}(1996)\citenamefont {Perdew},
  \citenamefont {Burke},\ and\ \citenamefont
  {Ernzerhof}}]{PhysRevLett.77.3865}%
  \BibitemOpen
  \bibfield  {author} {\bibinfo {author} {\bibfnamefont {J.~P.}\ \bibnamefont
  {Perdew}}, \bibinfo {author} {\bibfnamefont {K.}~\bibnamefont {Burke}}, \
  and\ \bibinfo {author} {\bibfnamefont {M.}~\bibnamefont {Ernzerhof}},\ }\href
  {\doibase 10.1103/PhysRevLett.77.3865} {\bibfield  {journal} {\bibinfo
  {journal} {Physical Review Letters}\ }\textbf {\bibinfo {volume} {77}},\
  \bibinfo {pages} {3865} (\bibinfo {year} {1996})}\BibitemShut {NoStop}%
\bibitem [{\citenamefont {Bl{\"{o}}chl}(1994)}]{PhysRevB.50.17953}%
  \BibitemOpen
  \bibfield  {author} {\bibinfo {author} {\bibfnamefont {P.~E.}\ \bibnamefont
  {Bl{\"{o}}chl}},\ }\href {\doibase 10.1103/PhysRevB.50.17953} {\bibfield
  {journal} {\bibinfo  {journal} {Physical Review B}\ }\textbf {\bibinfo
  {volume} {50}},\ \bibinfo {pages} {17953} (\bibinfo {year}
  {1994})}\BibitemShut {NoStop}%
\bibitem [{\citenamefont {Cococcioni}\ and\ \citenamefont
  {de~Gironcoli}(2005)}]{Cococcioni2005}%
  \BibitemOpen
  \bibfield  {author} {\bibinfo {author} {\bibfnamefont {M.}~\bibnamefont
  {Cococcioni}}\ and\ \bibinfo {author} {\bibfnamefont {S.}~\bibnamefont
  {de~Gironcoli}},\ }\href {\doibase 10.1103/PhysRevB.71.035105} {\bibfield
  {journal} {\bibinfo  {journal} {Physical Review B}\ }\textbf {\bibinfo
  {volume} {71}},\ \bibinfo {pages} {035105} (\bibinfo {year}
  {2005})}\BibitemShut {NoStop}%
\bibitem [{\citenamefont {Togo}\ and\ \citenamefont {Tanaka}(2015)}]{Togo2015}%
  \BibitemOpen
  \bibfield  {author} {\bibinfo {author} {\bibfnamefont {A.}~\bibnamefont
  {Togo}}\ and\ \bibinfo {author} {\bibfnamefont {I.}~\bibnamefont {Tanaka}},\
  }\href {\doibase 10.1016/j.scriptamat.2015.07.021} {\bibfield  {journal}
  {\bibinfo  {journal} {Scripta Materialia}\ }\textbf {\bibinfo {volume}
  {108}},\ \bibinfo {pages} {1} (\bibinfo {year} {2015})}\BibitemShut {NoStop}%
\bibitem [{\citenamefont {Martyna}\ \emph {et~al.}(1992)\citenamefont
  {Martyna}, \citenamefont {Klein},\ and\ \citenamefont
  {Tuckerman}}]{Martyna1992}%
  \BibitemOpen
  \bibfield  {author} {\bibinfo {author} {\bibfnamefont {G.~J.}\ \bibnamefont
  {Martyna}}, \bibinfo {author} {\bibfnamefont {M.~L.}\ \bibnamefont {Klein}},
  \ and\ \bibinfo {author} {\bibfnamefont {M.}~\bibnamefont {Tuckerman}},\
  }\href {\doibase 10.1063/1.463940} {\bibfield  {journal} {\bibinfo  {journal}
  {The Journal of Chemical Physics}\ }\textbf {\bibinfo {volume} {97}},\
  \bibinfo {pages} {2635} (\bibinfo {year} {1992})}\BibitemShut {NoStop}%
\bibitem [{\citenamefont {Liu}()}]{mcsolver}%
  \BibitemOpen
  \bibfield  {author} {\bibinfo {author} {\bibfnamefont {L.}~\bibnamefont
  {Liu}},\ }\href@noop {} {\bibinfo  {journal}
  {https://github.com/golddoushi/mcsolver {\rm (accessed Mar 25, 2021)}}\
  }\BibitemShut {NoStop}%
\bibitem [{\citenamefont {Liu}\ \emph {et~al.}(2020)\citenamefont {Liu},
  \citenamefont {Chen}, \citenamefont {Lin},\ and\ \citenamefont
  {Zhang}}]{Liu2020}%
  \BibitemOpen
\bibfield  {journal} {  }\bibfield  {author} {\bibinfo {author} {\bibfnamefont
  {L.}~\bibnamefont {Liu}}, \bibinfo {author} {\bibfnamefont {S.}~\bibnamefont
  {Chen}}, \bibinfo {author} {\bibfnamefont {Z.}~\bibnamefont {Lin}}, \ and\
  \bibinfo {author} {\bibfnamefont {X.}~\bibnamefont {Zhang}},\ }\href
  {\doibase 10.1021/acs.jpclett.0c01911} {\bibfield  {journal} {\bibinfo
  {journal} {Journal of Physical Chemistry Letters}\ }\textbf {\bibinfo
  {volume} {11}},\ \bibinfo {pages} {7893} (\bibinfo {year}
  {2020})}\BibitemShut {NoStop}%
\bibitem [{\citenamefont {Groenewolt}\ and\ \citenamefont
  {Antonietti}(2005)}]{Groenewolt2005}%
  \BibitemOpen
  \bibfield  {author} {\bibinfo {author} {\bibfnamefont {M.}~\bibnamefont
  {Groenewolt}}\ and\ \bibinfo {author} {\bibfnamefont {M.}~\bibnamefont
  {Antonietti}},\ }\href {\doibase 10.1002/adma.200401756} {\bibfield
  {journal} {\bibinfo  {journal} {Advanced Materials}\ }\textbf {\bibinfo
  {volume} {17}},\ \bibinfo {pages} {1789} (\bibinfo {year}
  {2005})}\BibitemShut {NoStop}%
\bibitem [{\citenamefont {Fei}\ \emph {et~al.}(2018)\citenamefont {Fei},
  \citenamefont {Dong}, \citenamefont {Feng}, \citenamefont {Allen},
  \citenamefont {Wan}, \citenamefont {Volosskiy}, \citenamefont {Li},
  \citenamefont {Zhao}, \citenamefont {Wang}, \citenamefont {Sun},
  \citenamefont {An}, \citenamefont {Chen}, \citenamefont {Guo}, \citenamefont
  {Lee}, \citenamefont {Chen}, \citenamefont {Shakir}, \citenamefont {Liu},
  \citenamefont {Hu}, \citenamefont {Li}, \citenamefont {Kirkland},
  \citenamefont {Duan},\ and\ \citenamefont {Huang}}]{Fei2018}%
  \BibitemOpen
  \bibfield  {author} {\bibinfo {author} {\bibfnamefont {H.}~\bibnamefont
  {Fei}}, \bibinfo {author} {\bibfnamefont {J.}~\bibnamefont {Dong}}, \bibinfo
  {author} {\bibfnamefont {Y.}~\bibnamefont {Feng}}, \bibinfo {author}
  {\bibfnamefont {C.~S.}\ \bibnamefont {Allen}}, \bibinfo {author}
  {\bibfnamefont {C.}~\bibnamefont {Wan}}, \bibinfo {author} {\bibfnamefont
  {B.}~\bibnamefont {Volosskiy}}, \bibinfo {author} {\bibfnamefont
  {M.}~\bibnamefont {Li}}, \bibinfo {author} {\bibfnamefont {Z.}~\bibnamefont
  {Zhao}}, \bibinfo {author} {\bibfnamefont {Y.}~\bibnamefont {Wang}}, \bibinfo
  {author} {\bibfnamefont {H.}~\bibnamefont {Sun}}, \bibinfo {author}
  {\bibfnamefont {P.}~\bibnamefont {An}}, \bibinfo {author} {\bibfnamefont
  {W.}~\bibnamefont {Chen}}, \bibinfo {author} {\bibfnamefont {Z.}~\bibnamefont
  {Guo}}, \bibinfo {author} {\bibfnamefont {C.}~\bibnamefont {Lee}}, \bibinfo
  {author} {\bibfnamefont {D.}~\bibnamefont {Chen}}, \bibinfo {author}
  {\bibfnamefont {I.}~\bibnamefont {Shakir}}, \bibinfo {author} {\bibfnamefont
  {M.}~\bibnamefont {Liu}}, \bibinfo {author} {\bibfnamefont {T.}~\bibnamefont
  {Hu}}, \bibinfo {author} {\bibfnamefont {Y.}~\bibnamefont {Li}}, \bibinfo
  {author} {\bibfnamefont {A.~I.}\ \bibnamefont {Kirkland}}, \bibinfo {author}
  {\bibfnamefont {X.}~\bibnamefont {Duan}}, \ and\ \bibinfo {author}
  {\bibfnamefont {Y.}~\bibnamefont {Huang}},\ }\href {\doibase
  10.1038/s41929-017-0008-y} {\bibfield  {journal} {\bibinfo  {journal} {Nature
  Catalysis}\ }\textbf {\bibinfo {volume} {1}},\ \bibinfo {pages} {63}
  (\bibinfo {year} {2018})}\BibitemShut {NoStop}%
\bibitem [{\citenamefont {Zhao}\ \emph {et~al.}(2019)\citenamefont {Zhao},
  \citenamefont {Zhang}, \citenamefont {Huang}, \citenamefont {Liu},
  \citenamefont {Zhang}, \citenamefont {He}, \citenamefont {Wu}, \citenamefont
  {Zhang}, \citenamefont {Wu}, \citenamefont {Yang}, \citenamefont {Gu},
  \citenamefont {Hu},\ and\ \citenamefont {Wan}}]{Zhao2019}%
  \BibitemOpen
  \bibfield  {author} {\bibinfo {author} {\bibfnamefont {L.}~\bibnamefont
  {Zhao}}, \bibinfo {author} {\bibfnamefont {Y.}~\bibnamefont {Zhang}},
  \bibinfo {author} {\bibfnamefont {L.-B.}\ \bibnamefont {Huang}}, \bibinfo
  {author} {\bibfnamefont {X.-Z.}\ \bibnamefont {Liu}}, \bibinfo {author}
  {\bibfnamefont {Q.-H.}\ \bibnamefont {Zhang}}, \bibinfo {author}
  {\bibfnamefont {C.}~\bibnamefont {He}}, \bibinfo {author} {\bibfnamefont
  {Z.-Y.}\ \bibnamefont {Wu}}, \bibinfo {author} {\bibfnamefont {L.-J.}\
  \bibnamefont {Zhang}}, \bibinfo {author} {\bibfnamefont {J.}~\bibnamefont
  {Wu}}, \bibinfo {author} {\bibfnamefont {W.}~\bibnamefont {Yang}}, \bibinfo
  {author} {\bibfnamefont {L.}~\bibnamefont {Gu}}, \bibinfo {author}
  {\bibfnamefont {J.-S.}\ \bibnamefont {Hu}}, \ and\ \bibinfo {author}
  {\bibfnamefont {L.-J.}\ \bibnamefont {Wan}},\ }\href {\doibase
  10.1038/s41467-019-09290-y} {\bibfield  {journal} {\bibinfo  {journal}
  {Nature Communications}\ }\textbf {\bibinfo {volume} {10}},\ \bibinfo {pages}
  {1278} (\bibinfo {year} {2019})}\BibitemShut {NoStop}%
\bibitem [{\citenamefont {Zhang}\ \emph {et~al.}(2020)\citenamefont {Zhang},
  \citenamefont {Zhou}, \citenamefont {Ge}, \citenamefont {Lin}, \citenamefont
  {Du}, \citenamefont {Zhong}, \citenamefont {Wang}, \citenamefont {Jiao},
  \citenamefont {Yuan}, \citenamefont {Tian}, \citenamefont {Chu},
  \citenamefont {Wu},\ and\ \citenamefont {Xie}}]{Zhang2020}%
  \BibitemOpen
  \bibfield  {author} {\bibinfo {author} {\bibfnamefont {N.}~\bibnamefont
  {Zhang}}, \bibinfo {author} {\bibfnamefont {T.}~\bibnamefont {Zhou}},
  \bibinfo {author} {\bibfnamefont {J.}~\bibnamefont {Ge}}, \bibinfo {author}
  {\bibfnamefont {Y.}~\bibnamefont {Lin}}, \bibinfo {author} {\bibfnamefont
  {Z.}~\bibnamefont {Du}}, \bibinfo {author} {\bibfnamefont {C.}~\bibnamefont
  {Zhong}}, \bibinfo {author} {\bibfnamefont {W.}~\bibnamefont {Wang}},
  \bibinfo {author} {\bibfnamefont {Q.}~\bibnamefont {Jiao}}, \bibinfo {author}
  {\bibfnamefont {R.}~\bibnamefont {Yuan}}, \bibinfo {author} {\bibfnamefont
  {Y.}~\bibnamefont {Tian}}, \bibinfo {author} {\bibfnamefont {W.}~\bibnamefont
  {Chu}}, \bibinfo {author} {\bibfnamefont {C.}~\bibnamefont {Wu}}, \ and\
  \bibinfo {author} {\bibfnamefont {Y.}~\bibnamefont {Xie}},\ }\href {\doibase
  10.1016/j.matt.2020.06.026} {\bibfield  {journal} {\bibinfo  {journal}
  {Matter}\ }\textbf {\bibinfo {volume} {3}},\ \bibinfo {pages} {509} (\bibinfo
  {year} {2020})}\BibitemShut {NoStop}%
\bibitem [{\citenamefont {Liu}\ \emph {et~al.}(2021{\natexlab{b}})\citenamefont
  {Liu}, \citenamefont {Zhang}, \citenamefont {Gao},\ and\ \citenamefont
  {Yan}}]{Liu2021}%
  \BibitemOpen
  \bibfield  {author} {\bibinfo {author} {\bibfnamefont {D.}~\bibnamefont
  {Liu}}, \bibinfo {author} {\bibfnamefont {S.}~\bibnamefont {Zhang}}, \bibinfo
  {author} {\bibfnamefont {M.}~\bibnamefont {Gao}}, \ and\ \bibinfo {author}
  {\bibfnamefont {X.-W.}\ \bibnamefont {Yan}},\ }\href {\doibase
  10.1103/PhysRevB.103.125407} {\bibfield  {journal} {\bibinfo  {journal}
  {Physical Review B}\ }\textbf {\bibinfo {volume} {103}},\ \bibinfo {pages}
  {125407} (\bibinfo {year} {2021}{\natexlab{b}})}\BibitemShut {NoStop}%
\bibitem [{\citenamefont {Liu}\ \emph {et~al.}(2021{\natexlab{c}})\citenamefont
  {Liu}, \citenamefont {Feng}, \citenamefont {Gao},\ and\ \citenamefont
  {Yan}}]{Liu2021a}%
  \BibitemOpen
  \bibfield  {author} {\bibinfo {author} {\bibfnamefont {D.}~\bibnamefont
  {Liu}}, \bibinfo {author} {\bibfnamefont {P.}~\bibnamefont {Feng}}, \bibinfo
  {author} {\bibfnamefont {M.}~\bibnamefont {Gao}}, \ and\ \bibinfo {author}
  {\bibfnamefont {X.-W.}\ \bibnamefont {Yan}},\ }\href {\doibase
  10.1103/PhysRevB.103.155411} {\bibfield  {journal} {\bibinfo  {journal}
  {Physical Review B}\ }\textbf {\bibinfo {volume} {103}},\ \bibinfo {pages}
  {155411} (\bibinfo {year} {2021}{\natexlab{c}})}\BibitemShut {NoStop}%
\bibitem [{\citenamefont {Goodenough}(1955)}]{Goodenough1955}%
  \BibitemOpen
  \bibfield  {author} {\bibinfo {author} {\bibfnamefont {J.~B.}\ \bibnamefont
  {Goodenough}},\ }\href {\doibase 10.1103/PhysRev.100.564} {\bibfield
  {journal} {\bibinfo  {journal} {Physical Review}\ }\textbf {\bibinfo {volume}
  {100}},\ \bibinfo {pages} {564} (\bibinfo {year} {1955})}\BibitemShut
  {NoStop}%
\bibitem [{\citenamefont {Anderson}(1959)}]{Anderson1959}%
  \BibitemOpen
  \bibfield  {author} {\bibinfo {author} {\bibfnamefont {P.~W.}\ \bibnamefont
  {Anderson}},\ }\href {\doibase 10.1103/PhysRev.115.2} {\bibfield  {journal}
  {\bibinfo  {journal} {Physical Review}\ }\textbf {\bibinfo {volume} {115}},\
  \bibinfo {pages} {2} (\bibinfo {year} {1959})}\BibitemShut {NoStop}%
\bibitem [{\citenamefont {Krukau}\ \emph {et~al.}(2006)\citenamefont {Krukau},
  \citenamefont {Vydrov}, \citenamefont {Izmaylov},\ and\ \citenamefont
  {Scuseria}}]{Krukau2006}%
  \BibitemOpen
  \bibfield  {author} {\bibinfo {author} {\bibfnamefont {A.~V.}\ \bibnamefont
  {Krukau}}, \bibinfo {author} {\bibfnamefont {O.~A.}\ \bibnamefont {Vydrov}},
  \bibinfo {author} {\bibfnamefont {A.~F.}\ \bibnamefont {Izmaylov}}, \ and\
  \bibinfo {author} {\bibfnamefont {G.~E.}\ \bibnamefont {Scuseria}},\ }\href
  {\doibase 10.1063/1.2404663} {\bibfield  {journal} {\bibinfo  {journal} {The
  Journal of Chemical Physics}\ }\textbf {\bibinfo {volume} {125}},\ \bibinfo
  {pages} {224106} (\bibinfo {year} {2006})}\BibitemShut {NoStop}%
\bibitem [{\citenamefont {Zhao}\ and\ \citenamefont {Wang}(2020)}]{Zhao2020}%
  \BibitemOpen
  \bibfield  {author} {\bibinfo {author} {\bibfnamefont {K.}~\bibnamefont
  {Zhao}}\ and\ \bibinfo {author} {\bibfnamefont {Q.}~\bibnamefont {Wang}},\
  }\href {\doibase 10.1016/j.apsusc.2019.144620} {\bibfield  {journal}
  {\bibinfo  {journal} {Applied Surface Science}\ }\textbf {\bibinfo {volume}
  {505}},\ \bibinfo {pages} {144620} (\bibinfo {year} {2020})}\BibitemShut
  {NoStop}%
\bibitem [{\citenamefont {Ma}\ \emph {et~al.}(2008)\citenamefont {Ma},
  \citenamefont {Lu},\ and\ \citenamefont {Xiang}}]{Ma2008}%
  \BibitemOpen
  \bibfield  {author} {\bibinfo {author} {\bibfnamefont {F.}~\bibnamefont
  {Ma}}, \bibinfo {author} {\bibfnamefont {Z.-Y.}\ \bibnamefont {Lu}}, \ and\
  \bibinfo {author} {\bibfnamefont {T.}~\bibnamefont {Xiang}},\ }\href
  {\doibase 10.1103/PhysRevB.78.224517} {\bibfield  {journal} {\bibinfo
  {journal} {Physical Review B}\ }\textbf {\bibinfo {volume} {78}},\ \bibinfo
  {pages} {224517} (\bibinfo {year} {2008})}\BibitemShut {NoStop}%
\bibitem [{\citenamefont {Liu}\ \emph {et~al.}(2016)\citenamefont {Liu},
  \citenamefont {Sun}, \citenamefont {Kawazoe},\ and\ \citenamefont
  {Jena}}]{Liu2016}%
  \BibitemOpen
  \bibfield  {author} {\bibinfo {author} {\bibfnamefont {J.}~\bibnamefont
  {Liu}}, \bibinfo {author} {\bibfnamefont {Q.}~\bibnamefont {Sun}}, \bibinfo
  {author} {\bibfnamefont {Y.}~\bibnamefont {Kawazoe}}, \ and\ \bibinfo
  {author} {\bibfnamefont {P.}~\bibnamefont {Jena}},\ }\href {\doibase
  10.1039/C5CP04835D} {\bibfield  {journal} {\bibinfo  {journal} {Physical
  Chemistry Chemical Physics}\ }\textbf {\bibinfo {volume} {18}},\ \bibinfo
  {pages} {8777} (\bibinfo {year} {2016})}\BibitemShut {NoStop}%
\bibitem [{\citenamefont {Zhang}\ \emph {et~al.}(2021)\citenamefont {Zhang},
  \citenamefont {Wang}, \citenamefont {Guo}, \citenamefont {Li},\ and\
  \citenamefont {Wang}}]{Zhang2021}%
  \BibitemOpen
  \bibfield  {author} {\bibinfo {author} {\bibfnamefont {Y.}~\bibnamefont
  {Zhang}}, \bibinfo {author} {\bibfnamefont {B.}~\bibnamefont {Wang}},
  \bibinfo {author} {\bibfnamefont {Y.}~\bibnamefont {Guo}}, \bibinfo {author}
  {\bibfnamefont {Q.}~\bibnamefont {Li}}, \ and\ \bibinfo {author}
  {\bibfnamefont {J.}~\bibnamefont {Wang}},\ }\href {\doibase
  10.1016/j.commatsci.2021.110638} {\bibfield  {journal} {\bibinfo  {journal}
  {Computational Materials Science}\ }\textbf {\bibinfo {volume} {197}},\
  \bibinfo {pages} {110638} (\bibinfo {year} {2021})}\BibitemShut {NoStop}%
\end{thebibliography}%

\end{document}